# Effects of magnetic field gradient and secondary electron emission on instabilities and transport in an E×B plasma configuration

M. Reza [*][†], F. Faraji[*], A. Knoll[*]

[*]Plasma Propulsion Laboratory, Department of Aeronautics, Imperial College London

[†]Corresponding Author (m.reza20@imperial.ac.uk)

**Abstract**: Today, partially magnetized low-temperature plasmas (LTP) in an E×B configuration, where the applied magnetic field is perpendicular to the self-consistent electric field, have important industrial applications. Hall thrusters, a type of electrostatic plasma propulsion, are one of the main LTP technologies whose advancement is hindered by the not-fully-understood underlying physics of operation, particularly, with respect to the plasma instabilities and the associated electron cross-field transport. The development of Hall thrusters with unconventional magnetic field topologies has imposed further questions regarding the instabilities' characteristics and the electrons' dynamics in these modern cross-field configurations. Accordingly, we present in this effort a series of studies on the influence of four factors on the plasma processes in the radial-azimuthal coordinates of a Hall thruster, namely, the magnetic field gradient, Secondary Electron Emission, electron-neutral collisions, and plasma number density. The studies are carried out using the reduced-order particle-in-cell (PIC) code developed by the authors. The setup of the radial-azimuthal simulations largely follows a well-defined benchmark case from the literature in which the magnetic field is oriented along the radius and a constant axial electric field is applied perpendicular to the simulation plane. The salient finding from our investigations is that, in the studied cases corresponding to elevated plasma densities, an inverse energy cascade leads to the formation of a long-wavelength, high-frequency azimuthal mode. Moreover, in the presence of strong magnetic field gradients, this mode is fully developed and induces a significant electron cross-field transport as well as a notable heating of the ion population.

## Section 1: Introduction

Low-temperature plasma technologies in which a partially magnetized plasma is subject to perpendicular electric and magnetic fields, such as the magnetrons for material processing and Hall thruster for spacecraft plasma propulsion, are today of high interest from both the applied and scientific perspectives. Particularly in case of the Hall thrusters, they are currently considered as optimal solutions for a variety of next-generation space mission scenarios due to their relatively higher propulsive metrics, easier manufacturing, and higher versatility of operation compared to other electrostatic electric propulsion solutions. Moreover, the rich underlying physics of plasma in Hall thrusters has made them a subject of active research in the last few decades. The E × B plasma configuration of these thrusters and the consequent large gradients and anisotropies in the plasma lead to a wide range of instabilities and oscillations developing across a broad spectrum of spatial and temporal scales. These processes are demonstrated to significantly affect the dynamics of the plasma species, in particular, the electrons [1]. The modification of the electrons' dynamics by the instabilities can affect the global performance and stability of the plasma discharge and, hence, the operation of the Hall thrusters. This intriguing coupling between the underlying physics and the operational behavior of Hall thrusters provides the academic research into the plasma phenomena in Hall thrusters an important applied significance since the resulting insights are a necessity for the development of high-performance, optimized devices and the reliable in-space application.

Establishing clear links between the plasma instabilities and the operation of Hall thrusters requires a comprehensive understanding of the nature and characteristics of the instabilities in each operating condition, characterizing the nonlinear interactions among the instability modes, and identifying the mechanisms by which these instabilities influence the electrons' dynamics, especially their cross-magnetic-field mobility, during various phases of their evolution and across the plasma conditions. In this regard, the three-dimensional, multiscale nature of the instabilities in Hall thrusters' plasma, which has become increasingly more evident by the mounting body of research [2], poses a major challenge for progress in this direction.

Nevertheless, in the absence of high-fidelity 3D simulations of the plasma in Hall thrusters, the investigation of the above aspects has been pursued in various 2D configurations and has led to valuable insights into dominant instability modes, their interplays, and their contributions to electrons' cross-field transport. In this regard, study of the instabilities and turbulence in the radial-azimuthal coordinates of a Hall thruster has received particular attention in the recent years to evaluate the lesser-known impact of the plasma-wall interactions and the radial plasma gradients on the characteristics and dynamics of the instabilities and the resulting electrons' transport.

Among the efforts dedicated to the study of the instabilities and electron transport in the radial-azimuthal Hall thruster configuration, the excitation and evolution of the Electron Cyclotron Drift Instability (ECDI) [3][4], the effect of the plasma-wall interactions on this instability [5][6], and the role of the ECDI in electron transport [4] have been the main subjects of research. In particular, the nonlinear evolution of the ECDI and the impact of the radial direction on the dynamics of azimuthal instabilities is studied in Ref. [3], in which the authors observed that the instabilities evolve toward larger azimuthal wavelengths and, at the later stages of the nonlinear evolution, the excitation of the so-called "Modified Two-Stream Instability (MTSI)" intensifies this inverse energy cascade. The evolution of the instabilities toward longer wavelength modes has been also evidenced in Ref. [7].

Regarding the impact of the MTSI on the plasma properties, the simulations of Refs. [3] and [7] showed that, because of its wavevector component along the magnetic field, the instability causes strong heating of the electrons in the radial direction. They also observed that the MTSI plays a significant role in the electrons' axial transport. In fact, in Ref. [7], the authors distinguished the contributions of the ECDI and the MTSI to transport using a Dynamic Mode Decomposition (DMD) analysis and concluded that the MTSI, when present, has the dominant contribution.

The interaction of the ECDI and the secondary electrons emitted from the walls together with their combined roles in "anomalous" cross-field transport of electrons have been studied in Ref. [5]. From their kinetic PIC simulations, the authors showed that, in the presence of high Secondary Electron Emission (SEE) rates, the "beam-plasma" instability is excited which induces very large axial electron currents. Moreover, in Ref. [6], the individual effect of the SEE on the sheath characteristics and the electrons' transport was investigated by performing a wide-range parametric study. The authors observed that increasing the SEE yield coefficient from 0 to 1 enhances the near-wall cross-field mobility of the electrons by a factor of 2 while decreasing the mobility in the bulk due to the cooling effect of secondary emitted electrons [6].

Finally, noting the recent development of Hall thrusters with the so-called "shielding" magnetic field topologies [8] designed to address the life-limiting erosion issue of the Hall thrusters' channel, particularly the high-power ones, which are intended for extended operational lifetimes, some limited research has been carried out to investigate the instabilities' characteristics and the electrons' transport in these unconventional field configurations. In this respect, the authors in Ref. [9] numerically solved the electrostatic dispersion relation of a homogenous, unbounded plasma with the conditions representative of the near-plume region of a magnetically shielded Hall thruster obtained from 2D axial-radial multifluid PIC simulations. They noticed the presence of the MTSI with the fastest growing mode occurring close to the front magnetic poles of the thruster [9]. As according to Ref. [10], the MTSI can heat up the ions perpendicular to the magnetic field almost to the same extent as it heats up the electrons parallel to field lines, the authors concluded that this could have a significant implication on the energy deposition on and sputtering of the magnetic poles [9].

Most recently, the impact of the magnetic field gradients and curvature on the plasma is analyzed using a 1D radial PIC simulations in Ref. [11]. In this publication, the authors performed simulations with various magnetic field curvatures and showed that the electron distribution function becomes more isotropic, and the tail of the electrons' distribution function is greatly replenished [11]. They also observed a concentrating effect in cases with concave magnetic field curvatures such as those encountered inside the channel of magnetically shielded Hall thrusters. It was noticed that the density in the center of the domain becomes higher than the case with a constant radial magnetic field [11].

Considering the above overview, the aim of the present article is to provide a comprehensive evaluation of the influence of various physical factors on the dynamics of the azimuthal instabilities and the consequent electrons' cross-field mobility using high-fidelity reduced-order kinetic simulations over relatively long timescales so as to cast light as well on the long-term evolution and interactions of the instabilities. The use of our reduced-order PIC code, introduced and verified in Refs. [12] and [13], serves as an essential enabler for this effort due to its remarkably lower computational cost compared to the conventional full-2D PIC codes and its high predictions' accuracy which is on a par with the traditional 2D codes. The factors whose influence we analyze in this paper are the magnetic field gradient, the Secondary Electron Emission, the electron-neutral collisions, and the plasma number density. As such, the analyses and discussions presented in this article are meant to expand the knowledge available in the literature concerning the radial-azimuthal physics of Hall thrusters and similar E × B discharges.

To perform the physical parametric studies in a simulation setup that is rigorously defined and is widely accepted within the E × B plasmas research community, we adopted, as the baseline setup, the one used for the radial-azimuthal kinetic code benchmarking activity [14]. Accordingly, the simulations in this work allow us to build upon the well-known results from the benchmark and to evaluate the variations in the observed behaviors due to each of the abovementioned four physical factors. Consequently, it is worth briefly describing the main physics that has been reported in the benchmark publication [14]. In this regard, the authors observed a cyclic quasi-steady evolution of the plasma discharge after an initial transient. During this transient, the ECDI and MTSI modes were developed. For detailed information about these instability modes, including the derivation and analysis of their dispersion relations, interested readers are referred to the existing literature, such as the Refs. [3][7][15][16][17][18] and the references therein. The nonlinear interactions between these instabilities were observed to lead to the periodic growth and damping of the two modes. When the MTSI modes were strongest, the radial electron temperature was seen to be at a local maximum. The damping of the MTSI resulted in a reduction of the radial electron temperature. When the radial electron temperature reached a local minimum, it was demonstrated that the ECDI becomes dominant [14]. It was also emphasized in Ref. [14] that the periodic oscillations in the radial electron temperature are reflected in the time evolution of the plasma density, which is due to the coupling that the imposed ionization source introduces between these two parameters in the benchmark's simulation case.

**Section 2: Overview of the reduced-order IPPL-Q2D PIC code**

IPPL-Q2D is a quasi-2D electrostatic explicit kinetic code based on the generalized reduced-order PIC scheme [12]. The reduced-order PIC is an innovative plasma simulation technique, devised by the authors, to tackle the computational cost issue of the conventional PIC schemes. It is predicated on a dimensionality-reduction approach for the decomposition of the multi-dimensional Poisson's equation into a system of coupled 1D ODEs. The reduced-order scheme also features a decomposition of the domain into multiple "regions" [12]. Within each region, a fine discretization of the simulation dimensions is carried out using custom 1D cells which enables resolving the variations of the plasma properties separately along each simulation coordinate [13]. Accordingly, for a 2D simulation case, the number of required computational cells and, hence, the total number of macroparticles can be reduced from $O(N^2)$ to $O(N)$, with $N$ being the number of cells along each simulation direction. This translates into significantly lower computational resource demand of the quasi-2D PIC compared to the traditional full-2D codes [12]. The detailed explanation of the reduced-order PIC scheme and its underlying dimensionality-reduction formulation can be found in Refs. [12] and [13].

The overall structure and the algorithmic implementations of IPPL-Q2D are introduced in Refs. [12][13]. The reduced-order quasi-2D code is verified extensively in our previous works against the well-defined axial-azimuthal [12][19] and radial-azimuthal [13] benchmark cases available in the literature. Particularly relevant to the present paper, the verifications of IPPL-Q2D in the radial-azimuthal simulation setup of the benchmark showed that a 50-region quasi-2D simulation, which offers a factor of 5 speed-up with respect to a full-2D simulation, provides high-fidelity predictions of the underlying physical processes and the plasma properties' distribution that compare well with the full-2D results [13]. Consequently, the simulations in this work are carried out using this approximation order of the 2D problem.

For the investigation of the effects of SEE and electron-neutral collisions, we have used, respectively, the wall-interaction and the Monte-Carlo Collisions (MCC) module of IPPL-Q2D. The wall-interaction module is described in Ref. [20], which also reports some of its verification results. In this work, we have used the linear SEE model from the wall-interaction module, which is further explained in Section 4.2. The MCC module is verified as result of the benchmarking of the 1D version of the code [19] against the Capacitively Coupled Discharge benchmark case by Turner et. al. [21]. For all simulations below that are relevant to the study of the collisions' effect, the cross-sections for the collisions between electrons and xenon neutrals are taken from the Biagi-v7.1 Dataset [22][23].

**Section 3: Description of the simulation setup and conditions**

The setup and conditions of the simulations performed in this effort are overall based on the 2D radial-azimuthal benchmark case by Villafana et. al. [14]. In any case, certain modifications are introduced in the baseline benchmark setup to enable the investigation of the effect(s) of each specific physical aspect mentioned in Section 1. We highlight these modifications in the following after providing an overview of the baseline setup.

The simulation domain is a 2D $x - z$ Cartesian plane, representative of a radial-azimuthal section of a Hall thruster. The $x$-coordinate is along the radial direction, and the $z$-coordinate represents the azimuthal direction. The y-axis is directed along the axial direction. The domain is 1.28-cm long along both the radial and azimuthal directions. All simulations are run for 30 $\mu s$. A constant axial electric field is applied in all simulation cases. Table 1 presents the values of the main computational and physical parameters used for the simulations.

| Parameter | Value [unit] |
|---|---|
| **Computational parameters** | |
| Domain length ($L_x = L_z$) | 1.28 [cm] |
| Virtual axial length ($L_y$) | 1 [cm] |
| Cell size ($\Delta x = \Delta z$) | 50 [μm] |
| Number of cells in each direction ($N_i = N_k$) | 256 |
| Time step ($ts$) | $1.5 \times 10^{-11}$ [s] |
| Total simulated duration ($t_{sim}$) | 30 [μs] |
| Initial number of macroparticles per cell ($N_{ppc}$) | 100 |
| **Physical parameters** | |
| Initial plasma density ($n_{i,0}$) | $5 \times 10^{16}$ [m$^{-3}$] |
| Neutral density ($n_n$) [*For collisional simulations*] | $1 \times 10^{19}$ [m$^{-3}$] |
| Initial electron temperature ($T_{e,0}$) | 10 [eV] |
| Initial ion temperature ($T_{i,0}$) | 0.5 [eV] |
| Axial electric field ($E_y$) | 10,000 [Vm$^{-1}$] |
| Radial magnetic field intensity at the mid radial plane ($B_x$) | 0.02 [T] |
| Electric potential at the walls ($\phi_w$) | 0 [V] |

Table 1: Summary of the computational and physical parameters used for the radial-azimuthal quasi-2D simulations

At the beginning of the simulations, the electrons and ions are sampled from a Maxwellian distribution at 10 eV and 0.5 eV, respectively, and are loaded uniformly throughout the domain at exactly the same positions. In order to limit the growth of the azimuthal waves and the particles' energy in the simulations [24], the approach of Ref. [14] is pursued, and a virtual axial extent with the length of $L_y = 1$ cm is considered. The particles crossing a domain's boundary along the axial direction are resampled from their initial Maxwellian distribution and are re-injected onto the simulation plane, maintaining their azimuthal and radial positions.

In all simulation cases, a zero-volt Dirichlet boundary condition is used for the electric potential at the walls. It is noteworthy that, in a Cartesian radial-azimuthal configuration, the symmetry of the simulation domain results in the absence of any net electron and ion current reaching the radial boundaries, which implies that the electrically floating boundary condition is automatically satisfied at the walls [6]. Hence, the adopted Dirichlet boundary condition in this study is reasonably well representative of the more realistic floating wall condition in modern Hall thrusters that feature a dielectric wall material. A periodic boundary condition is considered in the potential solver for the nodes at the two azimuthal ends of the domain. Moreover, particles that exit the domain along the azimuth from one side are re-introduced into the domain from the opposite side.

In order to compensate for the flux of particles lost to the walls, the approach of the radial-azimuthal benchmark [14] is adopted, and an ionization source is imposed. As described in more detail in Ref. [14], the ionization source is uniform in the azimuthal ($z$) direction and has a cosine distribution along the radius ($x$), extending from $x = 0.09\ cm$ to $x = 1.19\ cm$. The peak of the ionization source ($S_0$) is nominally $8.9 \times 10^{22}\ m^{-3}s^{-1}$, which corresponds to an axial ion current density of $100\ Am^{-2}$. The electron-ion pairs injected at each time step due to the ionization source are sampled from a Maxwellian at the respective initial temperature of each species.

We underline that adopting an approach to compensate for the radial flux of particles is a necessity for achieving steady-state condition in all simulations studying the plasma-wall interactions in 1D and 2D which do not self-consistently resolve the axial fluxes of the plasma species. In this regard, the adoption of an ionization source is one of the proposed approaches in the literature [4][14][25], which, as explained in Section 1, results in the simulations reaching a quasi-steady state.

Concerning the aspects of the simulation setup specific to each investigated case in Section 4, for the simulations in Section 4.1, which are dedicated to the study of the effects of the magnetic field gradients, the field intensity has a distribution along the radial coordinate with the magnitude at the mid radial plane being equal to 20 mT. In all other simulated cases, the magnetic field has only a constant radial component with the same magnitude of 20 mT.

In Section 4.2, which is related to the study of the influence of SEE and the electron-neutral collisions, the electrons' radial boundary condition is different with respect to the other simulations carried out. Indeed, whereas electrons hitting a wall are, in general, removed from the simulation, the Secondary Electron Emission is accounted for using a linear model for the simulations of Section 4.2. The secondary emitted electrons are sampled from a Maxwellian at the assumed temperature of 2 eV. Regarding the ions, they are always removed from the simulation in all studied cases if they cross a radial boundary.

The simulations are collisionless expect for those whose results are reported in Figure 17 to Figure 19 of Section 4.2. For these simulations, ionization, four excitations, and elastic momentum-exchange collisions are considered for the electron's interactions with the neutrals, which is resolved following the Monte-Carlo Collision algorithm [26]. For all collisions, a colliding electron undergoes an isotropic scattering. The electron's energy loss is also considered for the inelastic collisions. In case of an ionization collision, however, no new particles are injected.

Finally, in Section 4.3, where we study the effects of the plasma number density (or, equivalently, the axial current density), the nominal peak of the ionization source ($S_0$) is multiplied by a factor $\alpha$ corresponding to the ratio of the imposed current density for each case to the nominal one ($100\ Am^{-2}$). Accordingly, the initial plasma density ($n_{i,0}$) was scaled by the same factor, whereas the cells' size ($\Delta x$ and $\Delta z$) and the simulations' timestep were scaled by $1/\sqrt{\alpha}$.

**Section 4: Effects of radial magnetic field gradients, SEE, collisions, and current density on the radial-azimuthal plasma phenomena**

In this section, we present and discuss the results from our 50-region quasi-2D simulations aimed at studying the individual influence(s) on the radial-azimuthal dynamics of the plasma due to each of following four factors: (1) Radial gradients in the magnetic field, (2) Secondary Electron Emission, (3) electron-neutral collisions, and (4) plasma number density.

**4.1. The effect of radial gradients in the magnetic field**

Figure 1 shows the radial profiles of the radial ($B_x$) and axial ($B_y$) components of the magnetic field, as well as the radial distributions of the magnetic field intensity ($B$) and the angle ($\theta$) between the magnetic field lines and the radial coordinate ($x$) for three studied magnetic field configurations. Moreover, Figure 1(e) presents a 2D schematic of the adopted magnetic field topology in this work. Referring to plot (a) in Figure 1, among the three studied magnetic field configurations, Config. 1 corresponds to that of the baseline benchmark setup. Config. 2 has a convex radial profile of $B_x$, with the magnitude of the radial component increasing from $B_x = 0$ at the walls to the maximum value at the mid radial location. Config. 3 has instead a concave $B_x$ radial profile, with the maximum $B_x$ occurring at the two radial ends of the domain. Magnetic field configurations 2 and 3 are representative of the "magnetic-shielding" topology. In this respect, it is seen in Figure 1(e) that Config. 2 resembles the shielding configuration inside a Hall thruster's discharge channel, whereas Config. 3 corresponds to the field topology slightly outside the channel and within the "magnetic lens" [27].

From plot (c) in Figure 1, it is noticed that the gradient in the magnetic field intensity is larger for Config. 3, which is expected to result in a more pronounced magnetic mirror effect due to a larger $\nabla B$ force, i.e., the force due to the gradient in the magnetic field. In addition, plot (b) in Figure 1 shows that $B_y$ varies from –30 to 30 mT from one wall to the other in Configs. 1 and 2 and becomes zero at the mid radial plane. Finally, looking at plot (d) in Figure 1, it is evident that, in Config. 2, the magnetic field lines are parallel to the walls at the radial extremes of the domain, which, as will be demonstrated later, results in a notably reduced flux of particles to the walls.

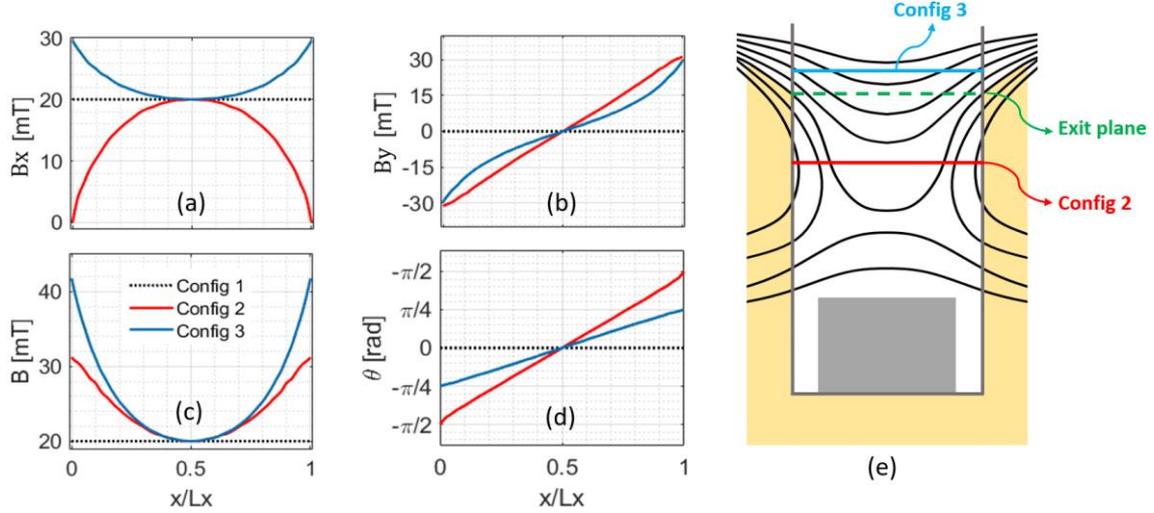

Figure 1: Radial profiles of the studied magnetic field configurations; (a) radial component ($B_x$), (b) axial component ($B_y$), (c) magnetic field intensity ($B$), and (d) the angle between the tangent to the magnetic field lines and the $x$-coordinate. In plot (e), a schematic of the adopted field topology and the cross-sections corresponding to Configs. 2 and 3 are shown.

We start presenting the results from various B-Config simulations with the radial distributions of the time-averaged plasma properties, ion number density ($n_i$), electron temperature ($T_e$), and ion temperature ($T_i$), in Figure 2. Looking at plot (a) in this figure, it is observed that, for Configs. 2 and 3, $n_i$ is much higher in the center of the domain compared to Config. 1. This higher density points to the confinement of the plasma by the magnetic field as the radial flux of plasma species is reduced in Configs. 2 and 3. In this respect, as the $\nabla B$ force is larger in Config. 3 (Figure 1(c)), the plasma is more concentrated in the center in this case compared to Config. 2. Similar observations are also reported in Ref. [11] from 1D radial simulations in which the effect of the magnetic field curvature on the radial distribution of the plasma properties was investigated.

Referring now to Figure 2(b), we notice that $T_e$ is relatively the same for Config. 2 compared to Config. 1, but it is much lower for Config. 3. As it will be shown in the following, the notably lower time-averaged $T_e$ for Config. 3 is due to the absence of the ECDI modes in this magnetic field configuration. Nonetheless, from plot (c) in Figure 2, it is seen that a significant ion heating has occurred in Config. 3. Indeed, the time-averaged $T_i$ profile for Config. 3 shows notably higher values across the radial extent of the domain. $T_i$ is also higher in Config. 2 compared to Config. 1, which implies some degree of ion heating in this configuration as well but clearly less significant than that in Config. 3.

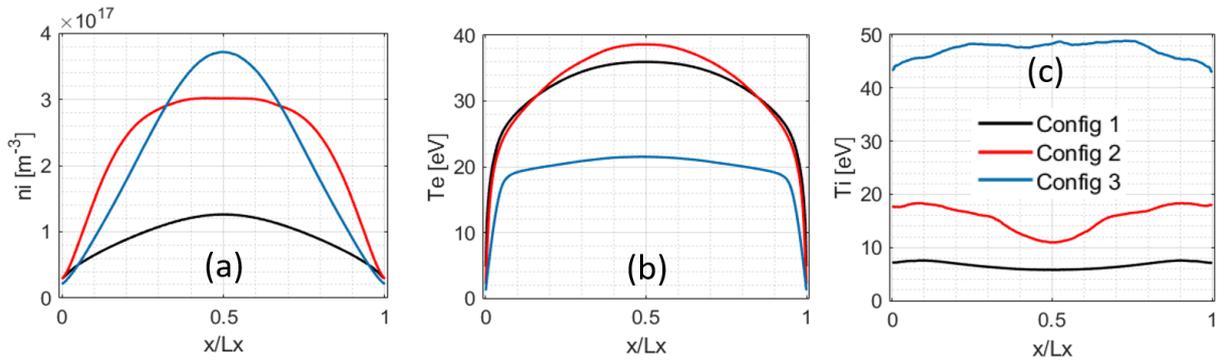

Figure 2: Radial profiles of plasma properties averaged over 25-30 $\mu s$ from the simulations with various magnetic field configurations; (a) ion number density, (b) electron temperature, (c) ion temperature.

Figure 3 shows the time evolution of the ion number density (plot (a)), electron temperature (plot (b)), average ions' kinetic energy (plot (c)), and the electrons' axial mobility (plot (d)), for the three magnetic field configurations. The ions' kinetic energy and the electrons' axial mobility are calculated using Eq. 1 and Eq. 2, respectively. In Eq. 1, $E_i$ is the ions' average kinetic energy in eV, $M_i$ is the ion mass, $e$ is the unit charge, $N_i$ is the total number of ion macroparticles, and $\boldsymbol{v}_{i,n}$ is the velocity vector of the n-th ion macroparticle. In Eq. 2, $\mu$ is the electrons' axial mobility, $v_{y_e}$ is the axial electron velocity, $E_y$ is the axial electric field, and $N_e$ is the total number of electron macroparticles.

$$E_i = \frac{1}{2}\left(\frac{M_i}{N_i e}\right)\sum_{n=1}^{N_i}(\boldsymbol{v}_{i,n} \cdot \boldsymbol{v}_{i,n}), \quad \text{(Eq. 1)}$$

$$\mu = \left|\frac{\sum_{n=1}^{N_e} v_{y_e}}{N_e E_y}\right|. \quad \text{(Eq. 2)}$$

Referring to Figure 3, it is observed that the fluctuations in various plasma parameters are of much higher amplitude and lower frequency for Config. 3 with respect to the other two magnetic field configurations. Moreover, the mean ions' kinetic energy and electrons' axial mobility is also higher for Config. 3 compared to Configs. 1 and 2. In Config. 2, the time evolution plots of ions' energy and electrons' mobility show higher-frequency, lower-amplitude oscillations and a lower time-averaged $E_i$ and $\mu$ compared to Config. 3. The mean $E_i$ and $\mu$ for Configs. 2 and 3 are both larger than the corresponding values for Config. 1.

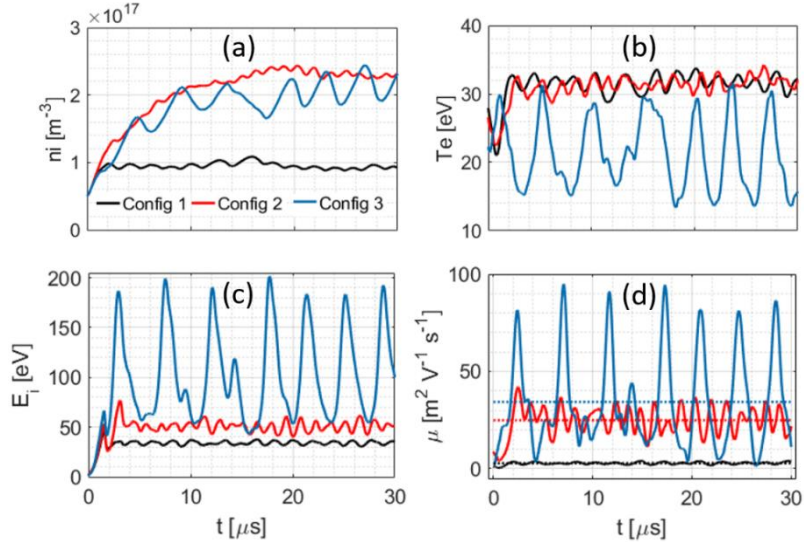

Figure 3: Time evolution of various plasma parameters from the simulations with various magnetic field configurations; (a) ion number density, (b) electron temperature, (c) average ions' kinetic energy, and (d) average electron's axial mobility. The horizontal dashed lines in plot (d) show the time-averaged mobility values.

The different temporal behaviors of the plasma parameters in Figure 3 suggest that, among the three magnetic configurations, there are notable variations in the underlying physical mechanisms and interactions. To illustrate this point, we first refer to Figure 4, in which we show, for each magnetic configuration, the 2D snapshots of various plasma properties at an instance of time corresponding to a local maximum in radial electron temperature ($T_{ex}$).

As described in Section 1, in the benchmark's simulation setup, which is identical to Config. 1, it has been observed that, when the radial electron temperature is maximum, the MTSI is fully formed and is dominant [13][14]. In this regard, the plots on the first row of Figure 4, which correspond to Config. 1, are reminiscent of the results from the benchmark [14] and our previous work [13], with the radial and azimuthal wavenumbers of the MTSI being particularly observable in the axial current density ($J_{ey}$) distribution. However, the 2D snapshots for Configs. 2 and 3 show notably different structures with respect to each other and also compared to Config. 1. Of course, in both configurations, MTSI-like patterns can be noticed, but the distributions and amplitudes are quite dissimilar.

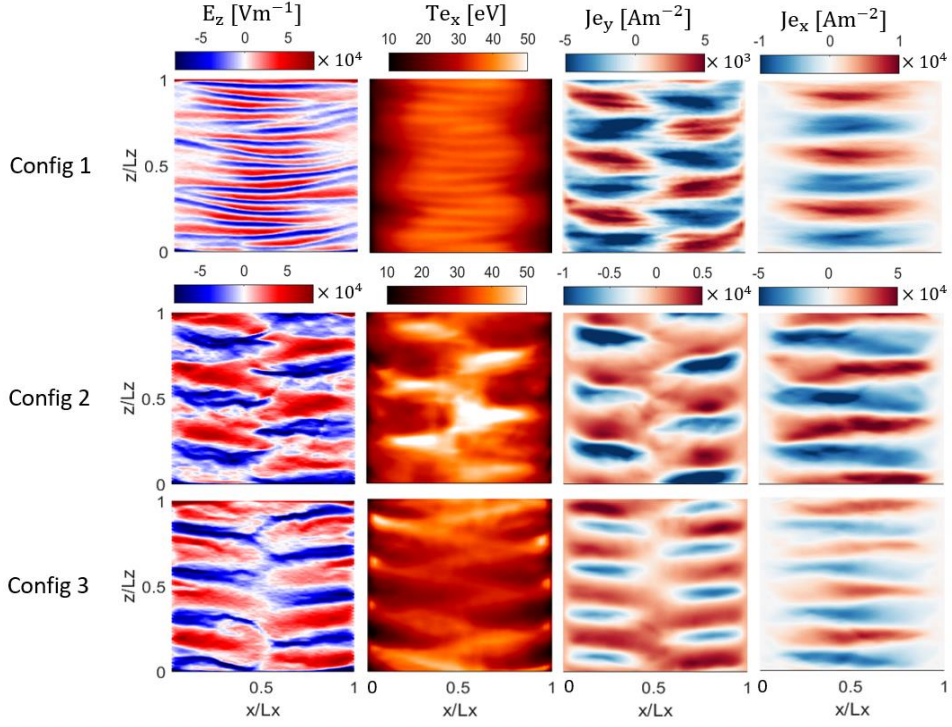

Figure 4: Comparison of the 2D snapshots of plasma properties at the time of local maximum of radial electron temperature for various magnetic field configurations. The columns, from left to right, represent the azimuthal electric field ($E_z$), radial electron temperature ($T_{ex}$), and axial and radial electron current densities ($J_{ey}$ and $J_{ex}$).

Second, in Figure 5, we have plotted for each magnetic configuration the 2D snapshots of the electron axial current density over one period of the discharge evolution ($T$). The video of the cyclic behavior shown in Figure 5 for the three B-Configs is available in Ref. [28], which more clearly visualizes the involved dynamics in each case. Looking at the evolution of $J_{ey}$ for Config. 1, the observed periodic behavior is similar to that reported in Refs. [13][14]. Indeed, at $t = t_0$, the MTSI is dominant. This instability starts to mitigate at $t_0 + \frac{T}{4}$, and the ECDI modes become eventually dominant at $t_0 + \frac{T}{2}$. The nonlinear interactions between MTSI and ECDI [14] leads to the disruption of the ECDI waves and the growth of the MTSI waves at $t_0 + \frac{3T}{4}$. Finally, the MTSI becomes fully formed and dominant again at $t = T$, and the cycle repeats.

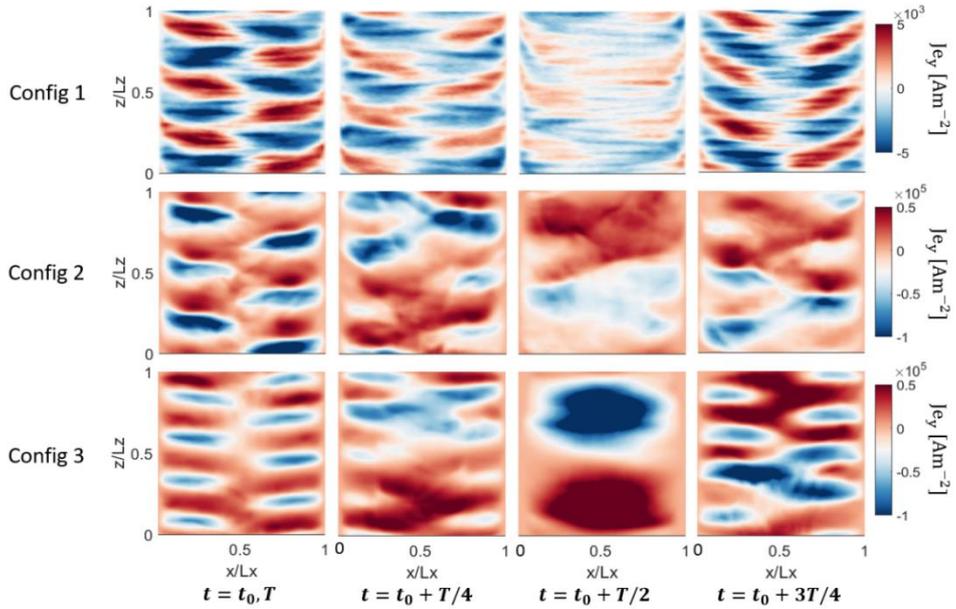

Figure 5: 2D snapshots of the axial electron current density at four different moments through the discharge evolution cycle for each magnetic field configuration. T represents the period of the cycle, which is different for each case.

In Configs. 2 and 3, the same cyclic growth and damping of the MTSI is visible. However, the damping of the MTSI modes in these two magnetic configurations seems to be associated with an inverse energy cascade that leads to the formation of a long-wavelength mode. This long-wavelength structure is more conspicuous in Config. 3, where the ion number density in the bulk plasma was higher and the electron temperature was lower compared to the other configurations (Figure 2). The large structure is then observed to break away from $t = t_0 + \frac{3T}{4}$, when the lower-wavelength MTSI-like modes start to grow and then become again dominant at $t = T$.

To further investigate the characteristics of the observed wave modes in Figure 4 and Figure 5, we have shown, in Figure 6, the spatiotemporally averaged 1D FFT plots of the azimuthal electric field signal from the B-Config simulations in three time intervals along the simulations. The plots in Figure 6 are obtained using the same approach detailed in Refs. [3][14]. The horizontal axis of the FFT plots is the normalized azimuthal wavenumber, $k_z/k_0$, where $k_0$ is the fundamental resonance wavenumber of the ECDI defined as $k_0 = \frac{\Omega_{ce}}{v_{de}}$ [14]. In the relation for $k_0$, $\Omega_{ce}$ is the electron cyclotron frequency, and $v_{de}$ is the electrons' azimuthal drift velocity. To calculate $\Omega_{ce}$ for Configs. 2 and 3, in which the magnetic field intensity varies along the radius, we have used the B value at the mid radial location, i.e., 20 mT, so that $k_0$ in these two cases is consistent with its value for Config. 1.

The FFT plots in Figure 6 confirm the observations made above regarding the 2D snapshots from various B-Config simulations. Indeed, we notice that, for Config. 1, the first and second harmonics of the ECDI at $\frac{k_z}{k_0} \sim 1$ and $\frac{k_z}{k_0} \sim 2$, respectively, as well as the MTSI mode at $\frac{k_z}{k_0} \sim 0.2$ are present in all time intervals. Nevertheless, for Configs. 2 and 3, the ECDI modes did not excite. For Config. 2, we notice the presence of the first and second harmonics of the MTSI, with the first harmonic occurring at the same $\frac{k_z}{k_0}$ as that for Config. 1. For Config. 3, the first MTSI harmonic is seen in all time intervals, but it is slightly shifted to shorter wavelengths compared to Configs. 1 and 2. The second MTSI harmonic for Config. 3 is only present during the time interval of 15-20 $\mu s$. Finally, for the magnetic configurations 2 and 3, we notice a long-wavelength mode at $\frac{k_z}{k_0} \sim 0.07$, which has been also observed in the corresponding 2D snapshots of Figure 5. For Config. 3, the long-$\lambda$ mode has the largest magnitude across the FFT spectrum.

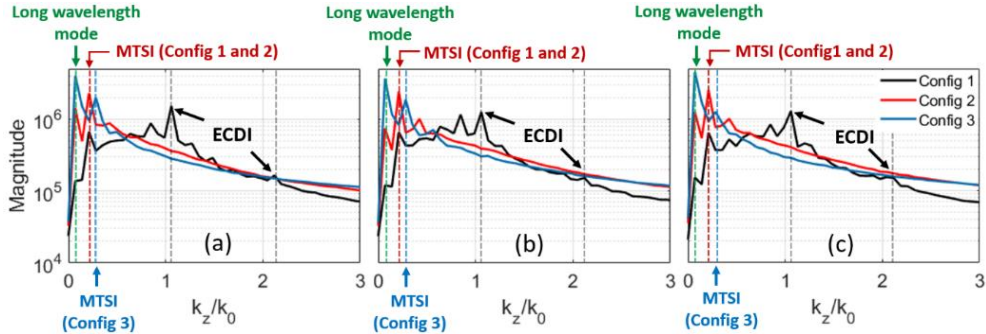

Figure 6: 1D FFT plots of the azimuthal electric field signal from various B-Config simulations, averaged over all azimuthal positions and over three time intervals; (a) 5-10 $\mu s$, (b) 15-20 $\mu s$, and (c) 25-30 $\mu s$.

We now assess the effect of the observed instability wave modes on the axial and cross-field mobility of the electrons. It should be noted that, in Config. 1, the magnetic field is radially constant and, thus, the axial and cross-field mobilities are equivalent. However, in Configs. 2 and 3, due to the radially varying magnetic field profile, the two mobility terms are different. In this regard, the time-averaged electrons' *axial* mobility, whose radial distribution for the three magnetic configurations is shown in Figure 7(a), is obtained using the time-averaged axial electrons' velocity ($v_{ey}$) and the (constant) axial electric field ($E_y$) as given by Eq. 3. The time-averaged electrons' *cross-field* mobility, whose radial profile is seen in Figure 7(b), is calculated using Eqs. 4 to 6 and according to the schematic illustrated in Figure 7(c).

$$\mu_y = \left|\frac{v_{ey}}{E_{ey}}\right|, \tag{Eq. 3}$$

$$v_{e\perp} = |v_{ey} \cos\theta| + |v_{ex} \sin\theta|, \tag{Eq. 4}$$

$$E_\perp = |E_y \cos\theta| + |E_x \sin\theta|, \tag{Eq. 5}$$

$$\mu_\perp = \frac{v_{e\perp}}{E_{e\perp}}. \tag{Eq. 6}$$

In Eqs. 4 to 6, $v_{e\perp}$ is the electrons' cross-field velocity, $v_{ex}$ is the electrons' radial velocity, $E_\perp$ is the magnitude of the electric field vector perpendicular to the magnetic field line at each radial location, $E_x$ is the radial electric field component, and $\mu_\perp$ is the cross-field mobility. $\theta$ is the angle between the x-axis and the tangent to the magnetic field line.

It is observed in Figure 7(a) that the radially averaged electrons' axial mobility is the highest for Config. 3, followed by that for Config. 2. The mean $\mu_y$ for Config. 1 is much lower than the corresponding values for Configs. 2 and 3. The mean values for the three configurations are expectedly consistent with those observed in Figure 3(d). The radial profile of $\mu_y$ for Config. 2 (Figure 7(a)) shows sharp increases toward the walls, which is due to the fact that the magnetic field lines in Config. 2 are along the y-direction near the walls (Figure 1(d)). Hence, it is also of interest to look at the electrons' cross-field mobility ($\mu_\perp$) in Figure 7(b). As it was expected, the rise in the mobility profile near the walls for Config. 2 is now less significant, and the mean $\mu_\perp$ for Config. 3 is more distinctly higher than that for Config. 2. Moreover, it is observed in plot (b) of Figure 7 that the cross-field mobility is overall lower in the center of the domain and increases toward the walls before decreasing again. This behavior is more pronounced for Config. 2 in which the time-averaged radial profile of $\mu_\perp$ also shows an oscillatory distribution, absent in the other two cases.

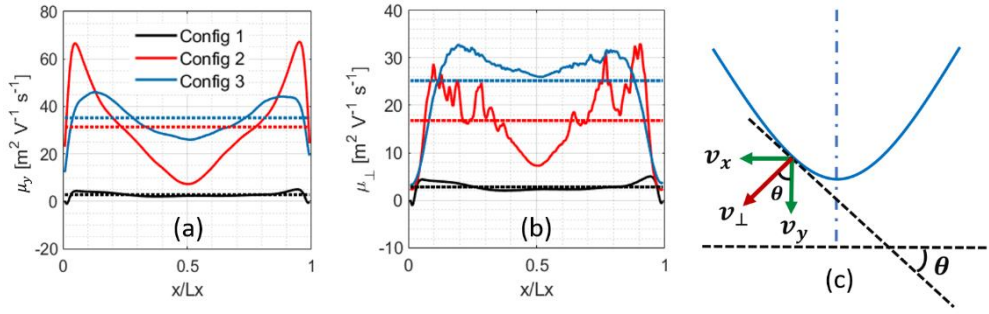

Figure 7: Comparison between the time-averaged (over 25-30 $\mu s$) radial profiles of (a) electrons' axial mobility, and (b) electrons' cross-field mobility for various magnetic field configurations. In (c), a schematic associated with the calculation of the cross-field velocity is shown; the blue curve represents the magnetic field line.

Considering the results shown in Figure 7 and noting that, for Config. 3, the long-$\lambda$ mode was seen to be dominant (Figure 5, third row, and Figure 6), it seems that the high cross-field mobility in this magnetic configuration is due to the long-$\lambda$ wave. To verify this speculation and to also identify the main contributor to the electrons' mobility among various instability modes in the other magnetic configurations, we present in Figure 8 the time evolution of the electrons' axial mobility and the amplitude of the dominant azimuthal wave modes ($|E_z|^2$) separately for Configs. 1 to 3 over the time interval of 5 to 30 $\mu s$.

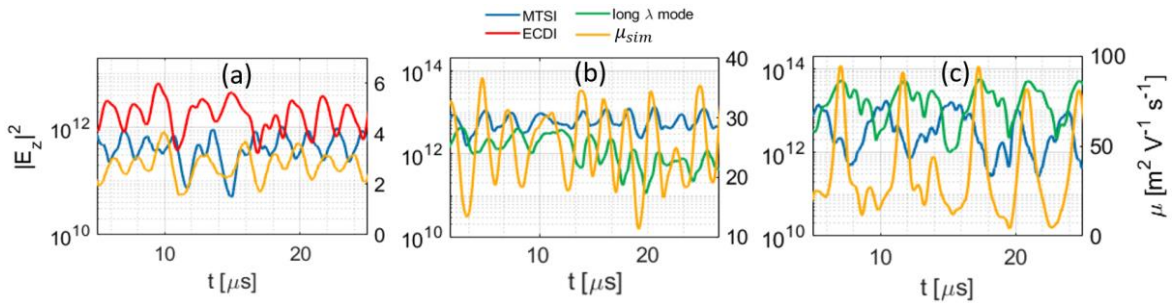

Figure 8: Time evolution of the amplitude of the dominant azimuthal modes (left axes) and the average electrons' mobility (right axes) for (a) Config. 1, (b) Config. 2, and (c) Config. 3. The approach pursued to plot the time evolution of $|E_z|^2$ for the dominant wave modes identified from the 1D FFTs (Figure 6) is explained in Ref. [14].

It is noticed that, for Config. 1 (Figure 8(a)), the time evolution of $\mu$ is in phase with the evolution of the amplitude of the dominant first harmonic of the ECDI, which underlines that the electrons' axial mobility in this case is mostly driven by the ECDI modes. For Config. 2 (Figure 8(b)), the fluctuations in $\mu$ are of larger amplitude compared to Config. 1 and are instead correlated with the oscillations in the amplitude of first MTSI harmonic, i.e., the dominant MTSI mode for Config. 2 as seen in Figure 6. Finally, for Config. 3, a rather violent temporal fluctuation in $\mu$ is observed in Figure 8(c), which is correlated with the oscillations in $|E_z|^2$ for the long-$\lambda$ mode. Indeed, the electrons' axial mobility is noticed to peak whenever the strong long-$\lambda$ instability becomes dominant.

Another interesting point from the plots in Figure 8 is that, for each configuration, the two dominant instability modes seem to interact and exchange energy such that when one mode strengthens, the other weakens. This behavior has been also reported in Ref. [14] for the interactions between ECDI and MTSI in the benchmark's setup, which is identical to Config. 1 (Figure 8(a)).

To conclude the discussions in this section, we look at the normalized radial and azimuthal velocity distribution functions for the electrons and the ions at the end of the simulations, i.e., at $t = 30\ \mu s$. The distribution functions are shown in Figure 9. It is observed in Figure 9(a) that, for Config. 1, the tail of the electrons' radial velocity distribution function is lost, whereas the confinement of the plasma in Configs. 2 and 3 has resulted in the retainment of the tail of the distribution functions. Moreover, the radial EVDF for Config. 2 spans across a broader range of velocities, which is consistent with the higher electron temperature observed for this configuration in Figure 2. Comparing the radial EVDF in plot (a) against the azimuthal one in plot (b) of Figure 9, the distribution functions for Configs. 2 and 3 seem to be rather isotropic, which is in line with the results of Ref. [11].

Concerning the radial and azimuthal ions' velocity distribution functions, we notice, from the radial IVDF (Figure 9(c)), that the distributions increasingly deviate from a Maxwellian from Config. 1 to Config. 3. Especially for Configs. 2 and 3, the IVDFs are depleted in the mid-velocity range, with the depletion being more significant for Config. 3. The azimuthal IVDFs for Configs. 2 and 3 (Figure 9(d)) show a considerable broadening, which is again more pronounced for Config. 3. The distortions of the IVDFs, particularly for Config. 3, suggest a strong interaction between the ions and the long-$\lambda$ waves, which seems to have caused a redistribution of the ions' energy from the radial direction to the azimuthal one.

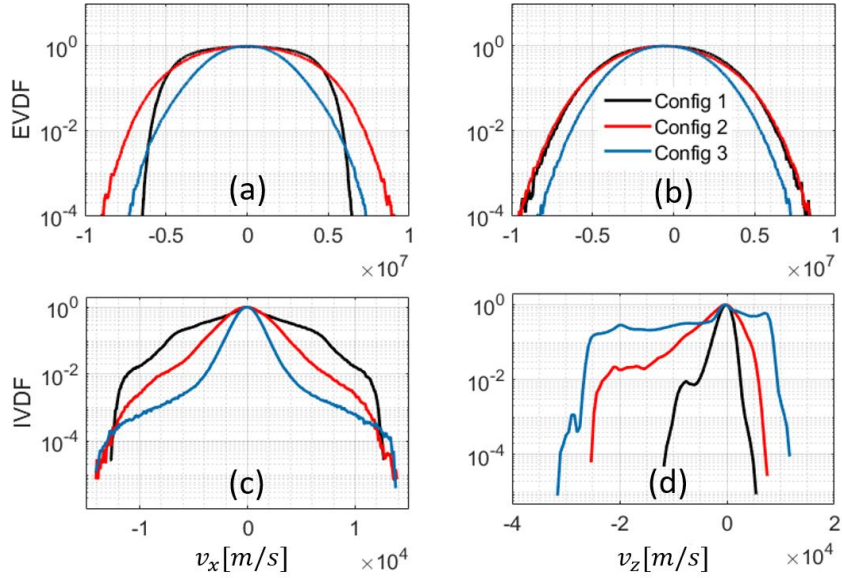

Figure 9: Normalized radial and azimuthal velocity distribution functions for various magnetic field configurations. First row: electrons' velocity distribution function along (a) the radial ($x$) and (b) the azimuthal ($y$) direction; second row: ions' velocity distribution function along (a) the radial and (b) the azimuthal direction.

### 4.2. The effect of Secondary Electron Emission and the electron-neutral collisions

In this section, we first present and discuss the results concerning the effect of the SEE phenomenon on the plasma processes along the radial-azimuthal coordinates. In this respect, the electron-neutral collisions are not considered in the following unless it is stated otherwise.

The SEE from the walls is accounted for using a linear model. A linear SEE model provides a rather simple estimate of the SEE yield coefficient ($\gamma$) according to Eq. 7.

$$\gamma(\epsilon) = \min\left(\gamma_{max}, \gamma_0 + \frac{\epsilon}{\epsilon^*}(1 - \gamma_0)\right). \tag{Eq. 7}$$

In the above equation, $\epsilon$ is the primary electron energy, $\gamma_{max}$ is the maximum electron emission coefficient, $\gamma_0$ is the probability of attachment, and $\epsilon^*$ is the crossover energy, i.e., the energy at which the SEE yield coefficient becomes equal to 1. For the studies presented in this article, we have kept the values of $\gamma_{max}$ and $\gamma_0$ constant and equal to their corresponding values for Boron Nitride, which are $\gamma_{max} = 2.9$ and $\gamma_0 = 0.578$. However, we have varied the $\epsilon^*$ value over a broad range, from 5 eV to 35 eV. Modifying the crossover energy values implies changing the slope of the linear part of the $\gamma$ vs. energy plot as it is illustrated in Figure 10.

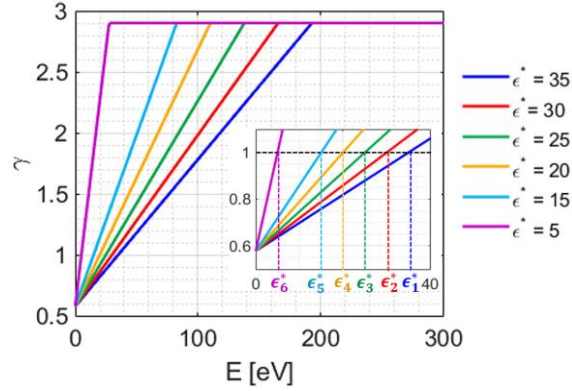

Figure 10: The variation of the secondary electron emission yield vs the electron energy for various $\epsilon^*$ values according to the linear re-remission model

From the plots (a) and (b) in Figure 11, it is noticed that decreasing the value of $\epsilon^*$ results in a consistent increase in the time-averaged ion number density and a decrease in the time-averaged electron temperature. In this regard, Figure 11(c) shows that the time-averaged SEE yield increases for lower $\epsilon^*$ values. Consequently, the higher emission rate of relatively cold secondary electrons reduces the $T_e$. This, in turn, translates into a smaller sheath potential drop for lower $\epsilon^*$, which reduces the radial flux of the ions, hence, increasing the ion number density in the bulk.

It is also interesting to note from Figure 11(c) that, for $\epsilon^* = 5$ eV, the time-averaged $\gamma$ is slightly below 1. As the $\epsilon^*$ is increased to 15 eV, $\gamma$ is stabilized at about the critical SEE yield value for xenon ($\gamma_{cr}$=0.985), which corresponds to a space-charge saturated sheath. Increasing $\epsilon^*$ further results in a monotonic decrease in $\gamma$.

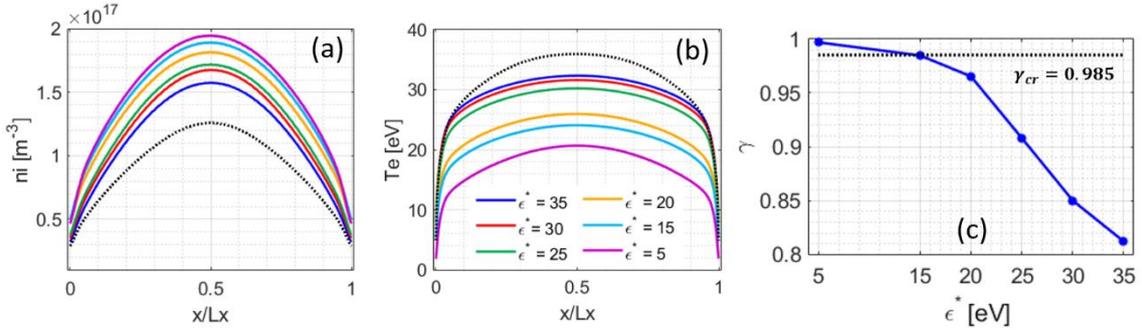

Figure 11: Time-averaged (over 25-30 $\mu s$) radial profiles of (a) ion number density, and (b) electron temperature for various $\epsilon^*$ values. Plot (c) shows the variation versus $\epsilon^*$ of the SEE yield, i.e., the ratio of the secondary emitted to primary electrons, averaged over the entire simulation duration.

Figure 12 presents the 2D snapshots of several plasma parameters from the simulations with various $\epsilon^*$ values at an instance of time corresponding to a local maximum in the radial electron temperature. The No-SEE case, i.e., the first column in Figure 12, corresponds to the benchmark's baseline setup and, as such, the 2D distributions in this case are representative of the results in Refs. [13][14], with the MTSI-like patterns being fully developed. However, as the SEE is taken into account and the value of $\epsilon^*$ is reduced from 25 eV to 5 eV, the overall structures in the 2D snapshots and the values of the plasma properties become notably different. In particular, an increasing cooling effect is clearly visible in the 2D snapshots of the radial electron temperature when the SEE rate becomes more and more significant. Furthermore, another interesting observation from the snapshots in Figure 12 is the dominance of a long-wavelength instability mode for $\epsilon^*$ values of 5 and 15 eV. In this respect, we observe from the plots of $E_z$ and, especially, $J_{ey}$ that, by decreasing the crossover energy, the 2D distributions change from that of the No-SEE case with dominant MTSI modes to those of the $\epsilon^* = 5$ eV where large-scale, coherent structures have appeared. A video of the cyclic evolution of the $J_{ey}$ from the No-SEE simulation case and the simulations with various $\epsilon^*$ values is available in Ref. [29].

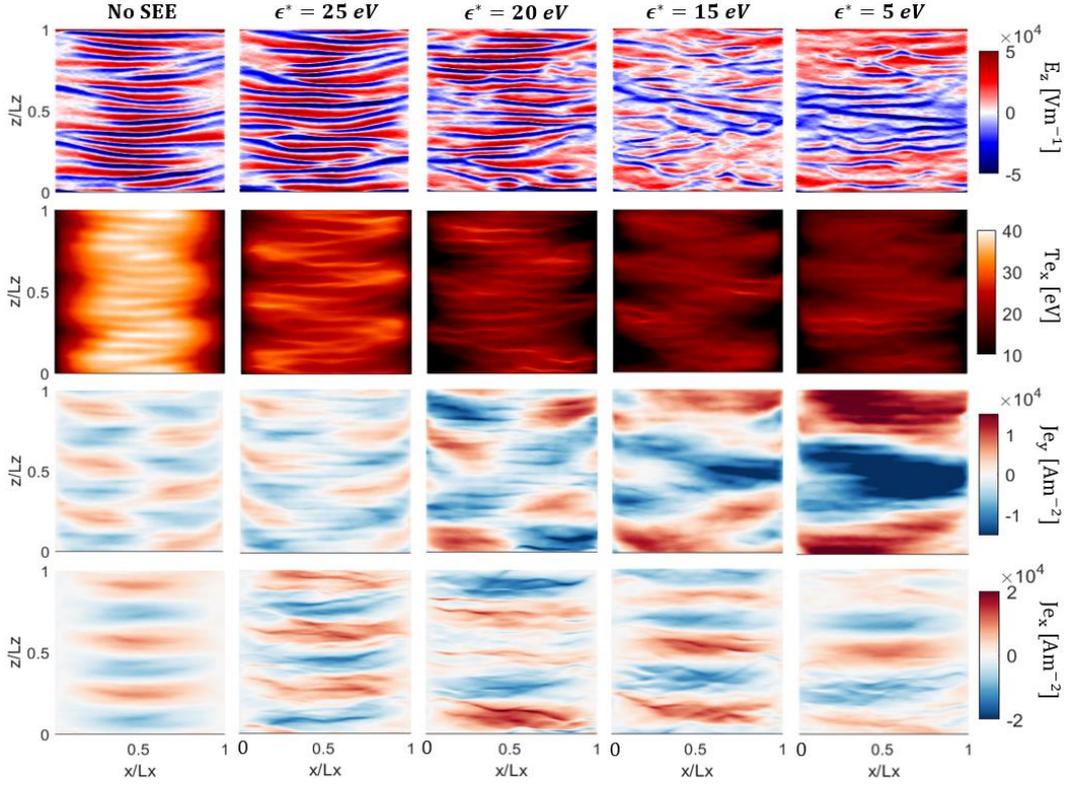

Figure 12: Comparison between the 2D snapshots of several plasma properties at the time of local maximum of radial electron temperature from the simulations with various $\epsilon^*$ against a simulation with no SEE. The rows, from top to bottom, represent the azimuthal electric field, radial electron temperature, and the axial and radial current densities.

The long-wavelength structures observed in Figure 12 for the lowest $\epsilon^*$ values (5 and 15 eV) are also noticeable in the spatiotemporal maps of the azimuthal electric field, plotted in Figure 13(b) and (c), as a characteristic large-scale modulation of the electric field signal at a frequency of about 1 MHz, which is absent for higher $\epsilon^*$ values, for instance $\epsilon^* = 20$ eV (Figure 13(a)).

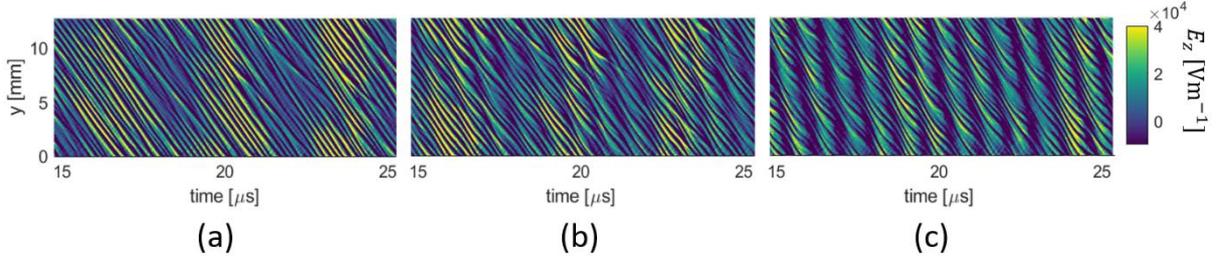

Figure 13: Spatiotemporal maps of the azimuthal electric field signal from three simulations with different $\epsilon^*$ values: (a) 20 eV, (b) 15 eV, and (c) 5 eV.

To analyze more closely the characteristics of the observed wave modes in Figure 12 and Figure 13, we refer to Figure 14, which shows the average 1D FFTs of the azimuthal electric field from the various-$\epsilon^*$ simulations in three time intervals along the simulation. From the plots (a) to (c) in Figure 14, it is evident that for $\epsilon^*$ values of 15 to 35 eV, the first harmonics of the ECDI and MTSI are clearly present throughout the simulations. However, for $\epsilon^*$ equal to 15 and 5 eV, we notice that the long-$\lambda$ wave is the dominant mode in all three time intervals. In addition, in case of $\epsilon^* = 5$ eV, the ECDI and MTSI modes are not distinctly visible, which is likely due to the cooling effect of the SEE population that has prevented the excitation and/or suppressed these two instabilities.

The suppression of the ECDI and MTSI modes for the simulation case of $\epsilon^* = 5$ eV is seen in Figure 15(a) to have resulted in a reduction in the electrons' axial mobility, both in the center of the domain ($\mu_{sim,center}$) and over the total radial extent ($\mu_{sim}$), contrary to the overall trend of the $\mu$ vs $\epsilon^*$ plot. In fact, it is observed in Figure 15(a) that, as the crossover energy is reduced, the total mobility and the mobility in the center of the domain increase.

This trend is also evident from Figure 15(b) which shows the time-averaged radial distribution of the electrons' mobility for various $\epsilon^*$ values. However, for $\epsilon^* = 5$ eV, whereas the difference between the $\mu_{sim,center}$ and $\mu_{sim}$, which implies the significance of the near-wall transport from the SEE, has increased compared to other cases, both mobility terms are lower than the corresponding values for $\epsilon^* = 15$ eV.

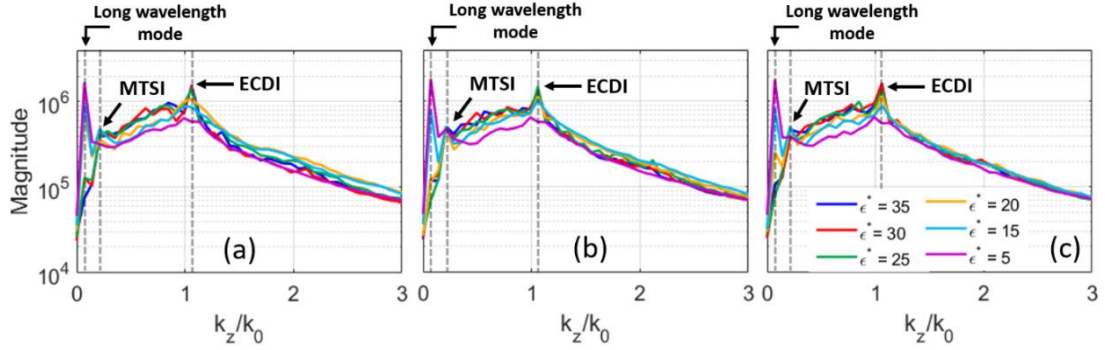

Figure 14: 1D FFT plots of the azimuthal electric field signal from the simulations with various $\epsilon^*$, averaged over all azimuthal positions and over the time intervals of (a) 5-10 $\mu s$, (b) 15-20 $\mu s$, and (c) 25-30 $\mu s$.

It is also worth pointing out that the increase in the near-wall mobility for decreasing values of crossover energy, observed in both plots (a) and (b) of Figure 15, is consistent with the previous results reported in the literature on the effect of the SEE on the electrons' transport [6].

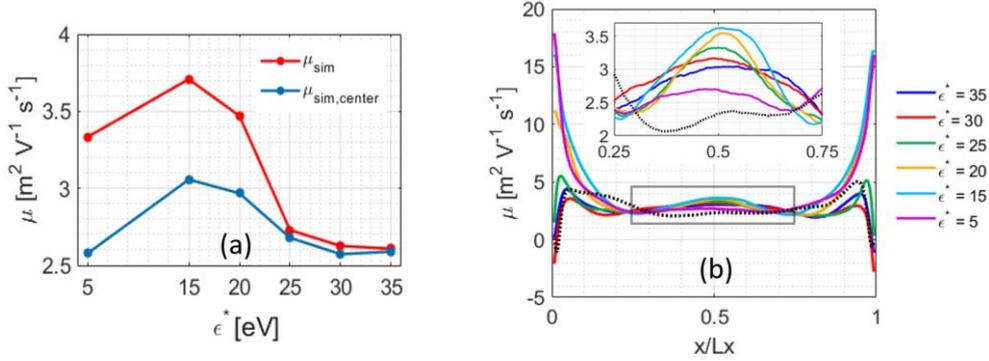

Figure 15: (a) Variation vs $\epsilon^*$ of the electrons' axial mobility terms, $\mu_{sim}$ and $\mu_{sim,center}$, averaged over the entire simulation domain and over time; (b) time-averaged (over 25-30 $\mu s$) radial profiles of the electrons' axial mobility for various $\epsilon^*$ values. The mobility terms $\mu_{sim}$ and $\mu_{sim,center}$ are both calculated using Eq. 2, but for $\mu_{sim,center}$, only electrons within the range of 0.25 to 0.75 $x/L_x$ are considered. The dashed black line in plot (b) corresponds to the No-SEE case.

As the final point on the SEE effects, we have plotted, in Figure 16, the normalized energy distribution function of the electrons from the simulations with various $\epsilon^*$ at $t = 30$ $\mu s$. It is evident from this figure that, as the SEE increases for lower crossover energies, the width of the distributions diminishes. This is because of the cooling effect of the secondary electrons, which was also reflected in the time-averaged profiles of the electron temperature (Figure 11(b)) for various values of $\epsilon^*$. Accordingly, we see in Figure 16 that the tail of the distributions become increasingly depleted from the No-SEE case to the case with $\epsilon^* = 5$ eV.

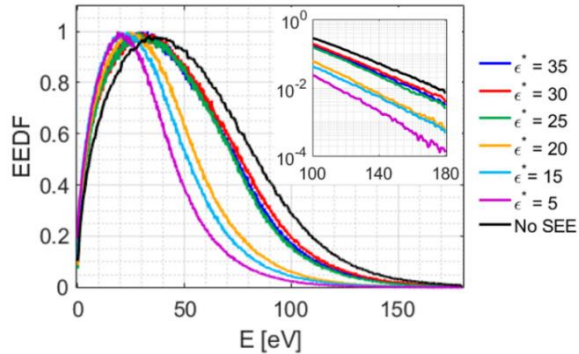

Figure 16: Normalized electrons' energy distribution function for various $\epsilon^*$ values. The zoomed-in view on the tail of the EEDF is also shown where the y-axis is in logarithmic scale.

We now look at the effect of the electron-neutral collisions on the plasma discharge in terms of the time-averaged distributions of the plasma properties, 1D FFT plots of the azimuthal electric field, and the time evolution of the electrons' mobility. Therefore, collisions are taken into account for the simulations whose results are presented in Figure 17 to Figure 19. The secondary electron emission is also considered for the simulations discussed here with two values of crossover energy, namely, $\epsilon^* = 15$ eV and $\epsilon^* = 35$ eV. In the first case, the SEE from the walls is strong whereas, in the second case, the SEE is relatively minor.

From the plots (a) and (b) in Figure 17, we notice that the collisions have reduced the time-averaged ion number density and the total electron temperature. The decrease in $T_e$ is, in part, due to the energy losses that the electrons will experience if they undergo inelastic collisions. The reason behind the decrease in the ion number density is, however, different between the cases with two $\epsilon^*$ values. In the case with $\epsilon^* = 35$ eV, it is observed from the FFT plots in Figure 18 that the collisions have caused an enhancement of the MTSI modes and a mitigation of the ECDI modes. As it is seen in Figure 17(c), the radial electron temperature ($T_{ex}$) is higher in the collisional case due to the heating of the electrons' population by the stronger MTSI modes. Consequently, the overall radial flux of particles reaching the wall increases, hence, reducing the time-averaged ion number density.

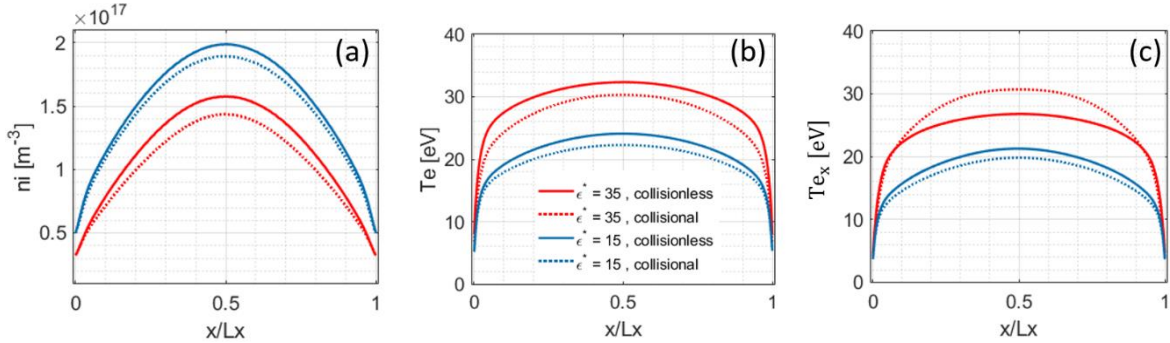

Figure 17: Comparison between the time-averaged (over 25-30 $\mu s$) radial profiles of (a) ion number density, (b) total electron temperature, and (c) radial electron temperature from the collisionless and collisional simulations with two different values of $\epsilon^*$.

In case of $\epsilon^* = 15$ eV, we observe in Figure 18 that the overall azimuthal wave content is not affected by the collisional processes, and the dominant instability modes are largely the same between the collisional and collisionless simulation. However, we showed in Figure 11(c) that, for $\epsilon^* = 15$ eV, the sheath is in the space-charge-saturated regime. In this regard, a possible explanation for the lower time-averaged ion number density in the collisional case compared to the collisionless one is that the collisions can disperse the electrons from the vicinity of the walls so that the sheath potential momentarily returns to its monotonic distribution near the walls. This behavior can on average yield an increased radial flux of the ions lost to the walls.

Regarding the effect of the collisional processes on the electrons' mobility, Figure 19(a) and (c) show that, for the $\epsilon^*$ value of 35 eV, whereas the mean $\mu_{sim}$ and $\mu_{sim,center}$ are the same in the absence of the collisions, the mean total mobility is larger in the collisional simulation. This shows that the electron-neutral collisions in this case have mostly enhanced the mobility of the electrons near the walls because the mean $\mu_{sim,center}$, which captures the mobility in the bulk plasma, is very similar between the collisional and collisionless simulation.

It is also observed in Figure 19(a) and (c) that the frequency of the fluctuations in the mobility terms has increased in the collisional case. In this regard, since we demonstrated in Section 4.1 that the oscillations in the mobility are correlated with the cyclic behavior of the discharge, corresponding to the excitation, growth, and damping of various azimuthal instabilities, we can conclude that the collisions have affected the timescale (period) of the nonlinear interactions between the azimuthal wave modes that underlie the cyclic discharge evolution.

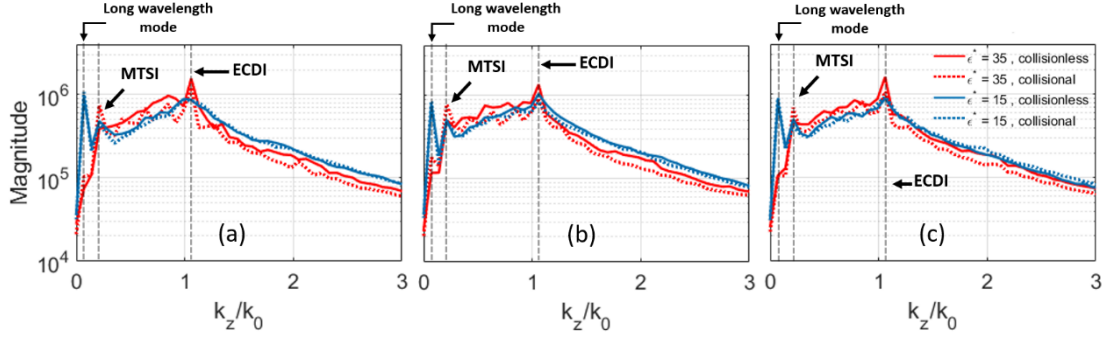

Figure 18: 1D FFT plots of the azimuthal electric field signal from the collisional and collisionless simulations, averaged over all azimuthal positions and over the time intervals of (a) 5-10 $\mu s$, (b) 15-20 $\mu s$, and (c) 25-30 $\mu s$.

For the case of $\epsilon^* = 15$ eV, we see, from Figure 19(b) and (d), that the collisional processes have not affected the mean values of the $\mu_{sim}$ and $\mu_{sim,center}$. This is due to the overall lower temperatures of the electrons near the walls in this case that lowers the probability of collision. Nevertheless, the presence of collisions in this case has led to the appearance of high-frequency features in the time evolution of the mobility terms. Investigation of the nature of these high-frequency fluctuations in mobility is left for future work.

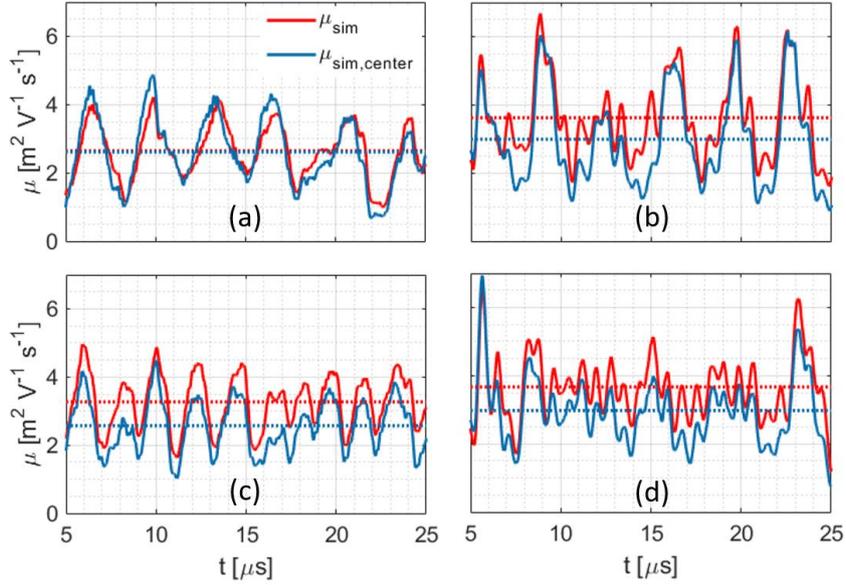

Figure 19: Time evolution of the electrons' axial mobility terms, $\mu_{sim}$ and $\mu_{sim,center}$, from the collisionless (first row) and the collisional simulations (second row): (a) and (c) correspond to $\epsilon^* = 35$ eV, whereas (b) and (d) are for $\epsilon^* = 15$ eV.

### 4.3. The effect of plasma number density

The last effect we have studied in this work is that of the plasma number density or, equivalently, the axial ion current density. Since the simulations feature an imposed ionization source, as described in Section 3, the plasma number density is controlled by the value of the peak of this source. As a result, to investigate the effect of plasma number density, we have changed the nominal peak of the ionization source, $S_0$, by a factor ranging from $\frac{1}{32}$ to 6.

In this regard, Figure 20 presents the radial distributions of the ion number density and the electron temperature for various peak values of the ionization source ($S$). Expectedly, as the peak of the ionization source is increased, so does the ion number density (Figure 20(a)). The increase in the ion (plasma) number density translates into stronger wave-particle interactions [19], which in turn causes a more significant heating of the plasma by the azimuthal instabilities and, thus, increasingly higher electron temperatures. This trend is clearly observed in the time-averaged electron temperature profiles in Figure 20(b).

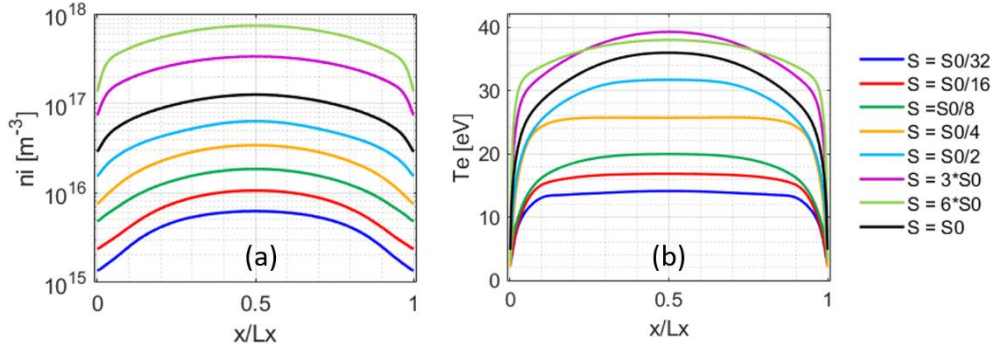

Figure 20: Time-averaged (over 25-30 $\mu s$) radial profiles of (a) ion number density, and (b) electron temperature from the simulations with various ionization source peak intensities ($S$).

The increase in the plasma number density also affects the azimuthal wave content of the discharge, and the overall associated magnitudes of the excited wave modes. This is illustrated in Figure 21, which shows the 1D FFT plots of the azimuthal electric field signal from the simulations with various values of $S$. From Figure 21, we can distinguish three distinct plasma regimes depending on the plasma number density, or the value of $S$.

First, from $\frac{S_0}{32}$ to $\frac{S_0}{8}$, the ECDI mode does not exist whereas the first harmonic of the MTSI is visible. The second harmonic of the MTSI is also excited for the $S = \frac{S_0}{16}$ and $\frac{S_0}{8}$. Second, for $S$ in the range of $\frac{S_0}{4}$ to $S_0$, the first and second harmonics of the MTSI, as well as the first harmonic of the ECDI are present. In the nominal case, the second ECDI harmonic at $\frac{k_z}{k_0}$ slightly higher than 2 is also visible. Third and last, for $3S_0$ and $6S_0$, the ECDI mode is not observed as a single mode but rather as a continuous spectrum around the wavenumber associated with this instability. Moreover, the long-$\lambda$ mode and the first harmonic of the MTSI are clearly visible. In the case of $S = 3S_0$, the second MTSI harmonic can be also seen in the time intervals of 5-10 $\mu s$ and 25-30 $\mu s$. The presence of the second MTSI harmonic in the case of $S = 3S_0$, which induces further heating of the electrons, can justify a slightly higher time-averaged $T_e$ for this case in Figure 20(b) compared to the case with $S = 6S_0$.

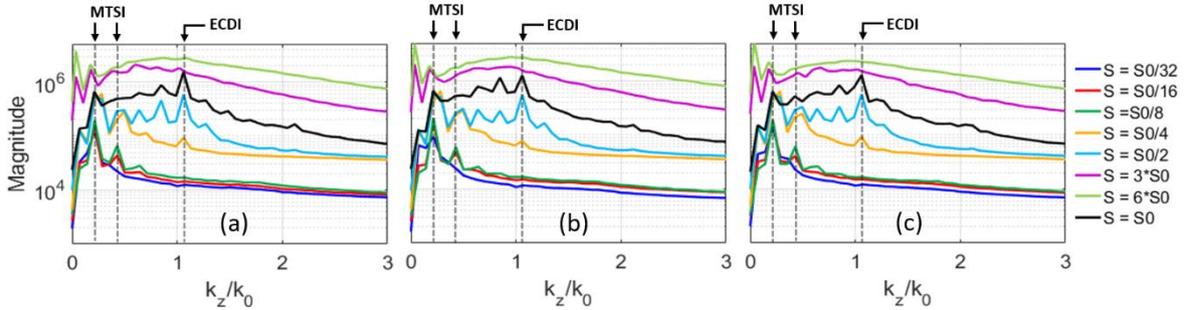

Figure 21: 1D FFT plots of the azimuthal electric field signal from the simulations with various $S$, averaged over all azimuthal positions and over the time intervals of (a) 5-10 $\mu s$, (b) 15-20 $\mu s$, and (c) 25-30 $\mu s$.

Moving on to the effect of plasma number density on the electrons' axial mobility, we observe, in Figure 22(a), that, as the plasma density increases, the spatiotemporally averaged electrons' mobility monotonically increases. This is an expected behavior since the FFT plots in Figure 21 illustrated that the overall intensity of the excited wave modes increases with the plasma number density and, additionally, the electron-wave interactions are more significant at a higher plasma density.

The trend mentioned above between the electrons' mobility and the plasma number density can be also noticed from the time-averaged radial distributions of the electrons' mobility in Figure 22(b). Referring to this figure, it is interesting to note that, for cases with $S$ in the range of $\frac{S_0}{32}$ to $\frac{S_0}{2}$, the mobility around the center of the domain, i.e., $\frac{x}{L_x} \sim 0.5$, is quite similar. However, as the spectrum of the excited azimuthal instabilities changes from the very low number density cases to those corresponding to $S$ values of $\frac{S_0}{4}$ and $\frac{S_0}{2}$, the mobility profiles show a more notable increase from the center of the domain toward the walls. For the nominal case with $S = S_0$, the mobility in the bulk is higher than the cases with $\frac{S_0}{4}$ and $\frac{S_0}{2}$, but it is lower near the walls. In the cases with $S$ value of $3S_0$ and $6S_0$, as the long-$\lambda$ mode becomes more dominant with increasing plasma density, the radial profile of the electrons' mobility overall moves toward larger values both in the bulk and near the walls.

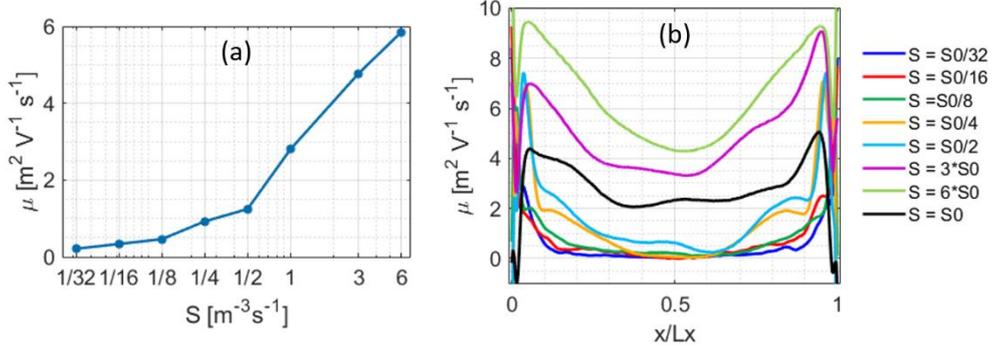

Figure 22: (a) Variation vs $S$ of the electrons' axial mobility averaged over the entire simulation domain and over time, (b) the time-averaged (over 25-30 $\mu s$) radial profiles of the electrons' axial mobility for various values of $S$.

The normalized velocity distribution functions of the ions and electrons along the radial and azimuthal directions are provided in Figure 23 for the simulations with various $S$ values at $t = 30\ \mu s$. From the plot (a) in Figure 23 corresponding to the electrons' radial VDF, we observe that, as the plasma density increases, the radial EVDF is increasingly broadened and moves toward a Maxwellian. The azimuthal EVDF (Figure 23(b)) shows an increasing heating of the electrons along the azimuth for higher plasma densities. These two observations are consistent with the fact that higher plasma number densities translate into stronger azimuthal waves and more significant interactions between the plasma species and the instabilities.

In this respect, the radial and azimuthal IVDFs, Figure 23(c) and (d), also exhibit clear signs of the heating of the ion population along both directions. This heating is most significant for the cases with $S = 3S_0$ and $6S_0$, where the long-$\lambda$ mode was seen to develop and become dominant.

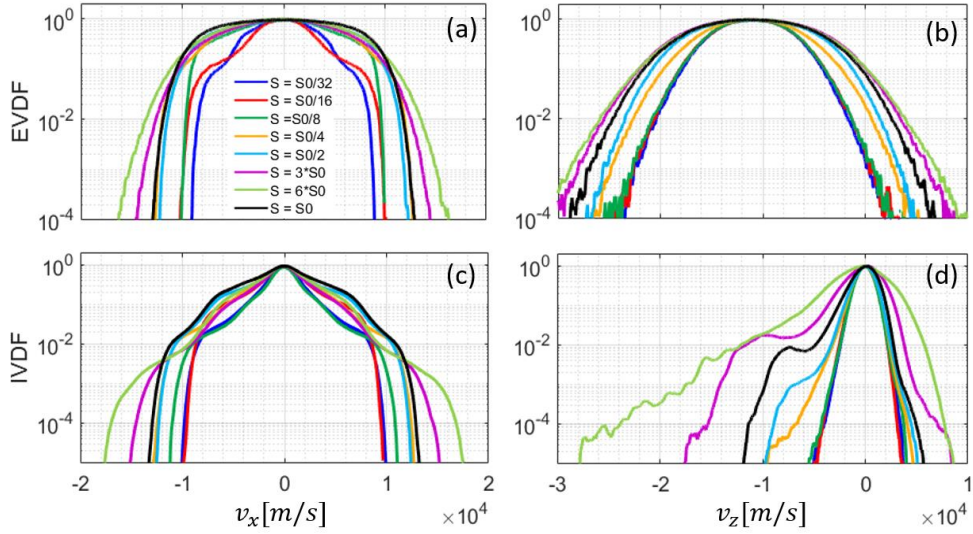

Figure 23: Normalized particles' velocity distribution functions for various values of $S$. First row: electrons' velocity distribution functions along (a) radial and (b) azimuthal direction; second row: ions' velocity distribution functions along (a) radial and (b) azimuthal direction.

Finally, in Figure 24, we have compared the 2D snapshots of various plasma properties through one evolution cycle of the discharge from the simulations corresponding to low and high plasma-number-density limits, namely, $S = \frac{S_0}{16}$ and $S = \frac{S_0}{8}$ vs $S = 3S_0$ and $S = 6S_0$. It is overall noticed that the visualizations in Figure 24 are consistent with the 1D FFT plots shown in Figure 21.

Looking more closely at the different panels in Figure 24 and the plots within, we notice that, in the cases with low plasma density, only MTSI-like structures are present. However, as the discharge evolves in one cycle, the first harmonic of the MTSI becomes dominant at $t = t_0 + \frac{T}{4}$, following by the dominance of the second harmonic at $t_0 + \frac{T}{2}$. The interactions between these MTSI harmonics causes a merging and splitting of the modes, leading to the tilted structures observable at $t = t_0$ and $t_0 + \frac{3T}{4}$.

In the high number density limit, on the contrary, the 2D plasma distributions are dominated by the long-wavelength mode superimposed on a spectrum of shorter wavelength ECDI and MTSI waves. The long-$\lambda$ wave mode is seen to be always dominantly present, but its amplitude slightly varies over the discharge's evolution cycle. A video of the dynamics of the discharge in terms of the time evolution of the axial electron current density ($J_{ey}$) from the simulations with various plasma number densities is available in Ref. [30].

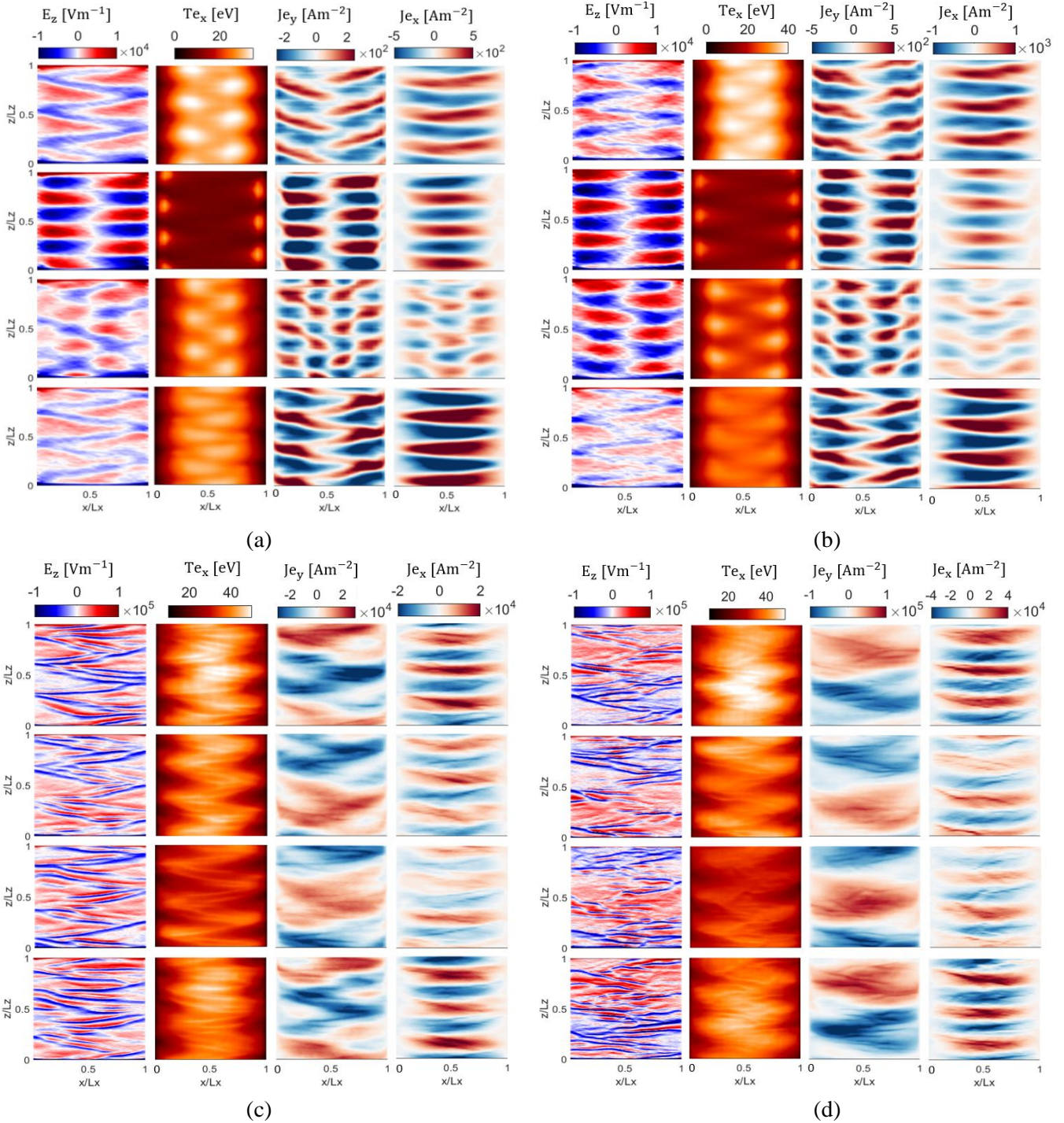

Figure 24: 2D snapshots of the plasma properties through one evolution cycle of the discharge from the simulations with (a) $S = S_0/16$, (b) $S = S_0/8$, (c) $S = 3S_0$, and (d) $S = 6S_0$. In each panel, the columns, from left to right, represent the azimuthal electric field, radial electron temperature, and the axial and radial electron current densities. Also, the rows, from top to bottom, represent various instances in one period ($T$) of the discharge evolution, i.e., $t_0$ and $T$, $t_0 + \frac{T}{4}$, $t_0 + \frac{T}{2}$, $t_0 + \frac{3T}{4}$.

The development of the long-$\lambda$ mode in the case of $S = 3S_0$ (Figure 24(c)) and the similar plasma number density in this case (Figure 20(a)) compared to the simulation in Section 4.2 with an SEE level corresponding to $\epsilon^* = 5$ eV (Figure 11(a)), for which the same long-wavelength mode was observed, allows us to link the observations and conclude that the formation of this mode is mainly driven by the plasma number density. Nevertheless, it is pointed out that the azimuthal wave content is case dependent and varies depending on the plasma conditions.

**Section 5: Conclusions**

In this article, we presented the results from an extensive parametric study in an E × B plasma configuration, representative of a radial-azimuthal section of a Hall thruster. The studies were focused on the effects of four physical factors on the characteristics and dynamics of the azimuthal instabilities and the consequent wave-induced electron transport in this 2D configuration. The numerous kinetic simulations whose results were discussed in this work were carried out using the reduced-order IPPL-Q2D PIC code at an order of approximation of the 2D problem corresponding to 50 regions along either the radial or azimuthal direction. The cost-effectiveness of the 50-region quasi-2D simulation, which was demonstrated in Ref. [13], and the interesting insights derived from this effort confirm the power of the reduced-order PIC code to be readily used as a tool for the verification of the theoretical predictions and/or to derive new theories to explain the nature of the physical phenomena of interest based on a more comprehensive set of observations regarding their variations over a wide range of plasma conditions. This is a desired capability that the significant computational cost of the traditional fully multi-dimensional PIC codes had so far hindered achieving.

The physical factors whose influence we investigated in this paper were the radial gradient in the magnetic field, the Secondary Electron Emission, the electron-neutral collisions, and the plasma number density. The common finding from all the simulations performed was that, in cases where the plasma density in the central part of the domain away from the walls is elevated due to any of the studied factors, an inverse energy cascade of the shorter wavelength instabilities, i.e., the MTSI and ECDI, seems to occur [3] that leads to the development of a long-wavelength, high-frequency wave mode. Even though the exact effect of this wave mode on the plasma species differs among the studied cases and depends on the conditions of the plasma, it overall causes a notable axial mobility of the electrons and a heating of the ion population.

To summarize the main effects observed from each of the specific physical factors, we demonstrated that the gradients in the magnetic field configuration affect the spectrum of the azimuthal instabilities, which consequently changes the dominant mechanism behind the electrons' axial mobility. In particular, for a simulation configuration representative of a cross-section of magnetically shielded Hall thrusters inside the discharge channel, we observed that the MTSI plays a dominant role in the electrons' transport. Whereas, for a configuration representative of the near-plume of these thrusters, the long-$\lambda$ mode was seen to play the major role in the electrons' axial mobility. To the best of our knowledge, this is the first time that the 2D radial-azimuthal physics of the Hall thrusters with a shielding magnetic field topology is being studied using high-fidelity kinetic simulations.

Concerning the effect of the SEE, we showed that decreasing the value of crossover energy leads to a consistent increase in the near-wall mobility. However, since the azimuthal wave content and the dominant instability modes were seen to vary as well with the strength of the SEE phenomenon, the variation of the overall average mobility vs $\epsilon^*$ did not show an always monotonic behavior and decreased for the lowest crossover energy value studied ($\epsilon^* = 5$ eV) due to the suppression of the ECDI and MTSI modes, leaving the long-$\lambda$ mode as the dominant one.

Regarding the effect of the collisions, the main observation was that the collisional processes alter the timescale of the nonlinear interactions between the azimuthal wave modes and, as such, the time evolution characteristics of the electrons' mobility changes between the collisional and collisionless cases.

Finally, we demonstrated that increasing the plasma number density translates into stronger azimuthal waves and more significant particles-wave interactions, leading to a monotonically increasing average electrons' axial mobility and an increasing heating of the ions along the radial and azimuthal directions. Moreover, increasing the plasma number density was shown to modify the azimuthal wave content. In this respect, for cases with the lowest plasma densities, only MTSI modes were seen to be present whereas, at the highest plasma densities, the long-$\lambda$ mode was observed to be dominant and coexist with a spectrum of shorter wavelength ECDI and MTSI waves.

**Acknowledgments**:

The present research is carried out within the framework of the project "Advanced Space Propulsion for Innovative Realization of space Exploration (ASPIRE)". ASPIRE has received funding from the European Union's Horizon 2020 Research and Innovation Programme under the Grant Agreement No. 101004366. The views expressed herein can in no way be taken as to reflect an official opinion of the Commission of the European Union.

**Conflict of Interest:**

The authors have no competing interests to declare that are relevant to the content of this article.

**Data Availability Statement**:

The data that support the findings of this study are available from the corresponding author upon reasonable request.